\documentclass[
reprint,
superscriptaddress,
amsmath,
amssymb,
aps,
pra
]{revtex4-2}

\usepackage{graphicx}% Include figure files
\usepackage{dcolumn}% Align table columns on decimal point
\usepackage{bm}% bold math
\usepackage[colorlinks, linkcolor=blue, anchorcolor=blue, citecolor=blue]{hyperref}
\usepackage{color}
\begin{document}

\preprint{APS/123-QED}

%\title{Entanglement Structure Detection via Hybrid CNN-Transformer Approach}% Force line breaks with \\
\title{Entanglement Structure Detection via Computer Vision}

%\thanks{A footnote to the article title}%
%\author{Ann Author}
% \altaffiliation[Also at ]{Physics Department, XYZ University.}%Lines break automatically or can be forced with \\
%\author{Second Author}%
% \email{Second.Author@institution.edu}
%\affiliation{%
% Authors' institution and/or address\\
% This line break forced with \textbackslash\textbackslash
%}%

\author{Rui Li}
\affiliation{School of Physics, Beihang University, Beijing 100191, China}

\author{Junling Du}
\affiliation{ School of Automation Science and Electrical Engineering, Beihang University, Beijing 100191, China}

\author{Zheng Qin}
\affiliation{School of Physics, Beihang University, Beijing 100191, China}

\author{Shikun Zhang}
\affiliation{School of Physics, Beihang University, Beijing 100191, China}

\author{Chunxiao Du}
\affiliation{School of Physics, Beihang University, Beijing 100191, China}

\author{Yang Zhou}
\email[Corresponding author: ]{yangzhou9103@buaa.edu.cn}
\affiliation{Research Institute for Frontier Science, Beihang University, Beijing 100191, China}

\author{Zhisong Xiao}
\affiliation{School of Physics, Beihang University, Beijing 100191, China}
\affiliation{School of Instrument Science and Opto-Electronics Engineering, Beijing Information Science and Technology University, Beijing 100192, China}

\date{\today}% It is always \today, today,
             %  but any date may be explicitly specified

\begin{abstract}
Quantum entanglement plays a pivotal role in various quantum information processing tasks. However, there still lacks a universal and effective way to detecting entanglement structures, especially for high-dimensional and multipartite quantum systems. Noticing the mathematical similarities between the common representations of many-body quantum states and the data structures of images, we are inspired to employ advanced computer vision technologies for data analysis. In this work, we propose a hybrid CNN-Transformer model for both the classification of GHZ and W states and the detection of various entanglement structures. By leveraging the feature extraction capabilities of CNNs and the powerful modeling abilities of Transformers, we can not only effectively reduce the time and computational resources required for the training process but also obtain high detection accuracies. Through numerical simulation and physical verification, it is confirmed that our hybrid model is more effective than traditional techniques and thus offers a powerful tool for independent detection of multipartite entanglement.

\end{abstract}

%Use showkeys class option if keyword
                              %display desired
\maketitle

%\tableofcontents

\section{\label{sec:level1}Introduction}

Quantum entanglement underlies the unique characteristics and advantages of quantum systems over classical systems \cite{PhysRev.47.777,RevModPhys.81.865,chitambar2019quantum}. It plays a vital role in various branches of quantum information \cite{bennett2000quantum,jaeger2007quantum} such as quantum communication \cite{RevModPhys.82.665}, quantum cryptography \cite{PhysRevLett.67.661}, and quantum computing 
 \cite{RevModPhys.74.347,nielsen2002quantum}. As a result,  knowledge about the structure of entangled states \cite{PhysRevX.8.021072, RevModPhys.81.865} is necessary, which is to say, it is essential to determine the entanglement type and entanglement depth \cite{PhysRevLett.67.661} of these states. For low-dimensional systems, there exists a necessary and sufficient condition for the separability of bipartite systems the Peres-Horodecki criterion based on the positivity of partial transpose (PPT) criterion \cite{PhysRevLett.77.1413,PhysRevA.98.012315}. However, for higher-dimensional systems, or more parties, there is no single universal separability condition. Various theoretical and experimental tools have been proposed to analyze and detect multipartite quantum entanglement. One of the most common approaches is entanglement witness (EW) \cite{horodecki1996teleportation,terhal2000bell,lewenstein2000optimization,guhne2009entanglement} which can distinguish a specific entangled state from separable ones.

However, existing entanglement detection methods lack universality \cite{guhne2009entanglement}. For example, one cannot determine the entanglement witness operator for all possible entangled states and an arbitrary number of partitions. Even worse is the lack of a general description of all entangled states in multipartite quantum systems. The biggest challenge for the analytic or numeric approaches is that as the qubit number increases, both the number of possible quantum states and the number of entanglement structures increases exponentially. Consequently, researchers have turned to machine learning techniques for help.

At first, most related studies adopted supervised training methods \cite{chen2021entanglement, greenwood2023machine,chen2022certifying,qiu2019detecting}, achieving high accuracy in specific tasks. However, the process of labeling a large number of quantum states is very time-consuming and almost impossible for multi-qubit systems. Thus, semi-supervised \cite{ zhang2023entanglement, luo2023detecting} and unsupervised learning \cite{chen2021detecting} techniques that can predict a large number of unlabeled quantum states from a small number of labeled states have been developed. The problem with these methods is they can only offer limited precision and require further experimental validation. Considering the structural resemblance between the mathematical representation of quantum states and classical image data, we are inspired to utilize established computer vision techniques for entanglement structure detection, which are known for their reduced resource consumption and computational efficiency.

In this study, a hybrid CNN-Transform model is proposed for the detection of multipartite entanglement. This network structure was initially developed for image processing and is particularly adept at handling large quantities of data. Thereinto, a deep convolutional neural network (CNN) \cite{gu2018recent,vedaldi2015matconvnet} can effectively identify local features and alter data dimensions through convolution operations while the self-attention mechanism of the Transformer \cite{liu2021swin,vaswani2017attention,dosovitskiy2020image} can capture long-distance dependencies. Thus, the trained neural network can identify the key features in different descriptions of entanglement structures, precisely delineating the boundaries between completely independent samples and various entangled samples. This brings about a major advantage: the same neural network can be applied to both the classification of Greenberger-Horne-Zeilinger (GHZ) \cite{zukowski1998quest} or W states \cite{eibl2004experimental} and the detection of specific entanglement structures. As far as we know, the current machine learning method can only handle a specific single task, such as determining whether a particular type of entanglement exists or detecting the separability and depth of some special multipartite entangled states. Different from them, our hybrid model demonstrates exceptional performance in accomplishing both classification and detection tasks with low time and computational cost. Numerical examples, ranging from three to ten qubit systems, demonstrate that our network achieves an average classification accuracy exceeding 99.57\% and a 95\% accuracy in detecting entanglement structures.

The structure of this paper is organized as follows. In Section \hyperref[sec:levelM]{II}, we give a brief introduction to the basic concepts and then propose the hybrid CNN-Transformer model to detect entangled structures in multi-qubit systems. In Section \hyperref[sec:level3]{III}, we quantify the performance of our trained neural network by numerical examples and real quantum devices. Finally, this paper is concluded in Section \hyperref[sec:level4]{IV}.

\section{\label{sec:levelM}Methods}

In this Section, We introduce the definition of GHZ and W entangled states used for data generation and the concept of entanglement depth used as a descriptor of the entanglement structure. Then, we describe the process of dataset preparation for our analysis and introduce a novel hybrid CNN-Transformer model that captures the inherent properties of quantum states to classify and detect entangled states.

\subsection{\label{sec:level2}GHZ and W Entanglement States}

In quantum mechanics, the pure states of quantum systems $A$ and $B$ are represented by thr state vectors $\left.|{\mathrm{\Psi}}_A\right\rangle$ and $\left.|{\mathrm{\Psi}}_B\right\rangle$ in their Hilbert spaces $H_A$ and $H_B$. The composite system $(A+B)$ is represented by the state vector $\left.|{\mathrm{\Psi}}_{AB}\right\rangle= \left.|{\mathrm{\Psi}}_A\right\rangle\otimes\left.|{\mathrm{\Psi}}_B\right\rangle$ in the Hilbert space $H_A\otimes H_B$. If $\left.|{\mathrm{\Psi}}_{AB}\right\rangle$ can be written as a product state, it is not entangled; otherwise, it is an entangled state \cite{nielsen2002quantum}. Generalized to an $N$-qubit case, its quantum state can be expressed as:
\begin{equation}
    \left.|{\mathrm{\Psi}}_{ABC\ldots}\right\rangle = \left.|{\mathrm{\Psi}}_A\right\rangle\otimes\left.|{\mathrm{\Psi}}_B\right\rangle\otimes\ldots\otimes\left.|{\mathrm{\Psi}}_N\right\rangle
    \label{eq:psi}
\end{equation}

If a quantum system cannot be described in terms of a pure state, then such a quantum state is called a mixed state and is usually represented by a density matrix:

\begin{equation}
    \hat{\rho} = \sum_i p_i\hat{\rho}_i,
    \label{eq:rho}
\end{equation}
where $p_i > 0$ and $\sum_i p_i = 1$.Generally, for an $N$-partite quantum state, if its density matrix $\rho$ can be expressed as a convex combination of multiple product states, it can be represented by the following equation:

\begin{equation}
    \hat{\rho} = \sum_l p_i\hat{\rho}_i^1\otimes\hat{\rho}_i^2\ldots\otimes\hat{\rho}_i^N
    \label{eq:rho_general}
\end{equation}

{\noindent}Here, $p_i$ is probability that $0\leq p_i\leq 1$ and $\sum_i p_i=1$, $\hat{\rho}_i^k $ is the pure state density matrix of each subsystem. If the density matrix can be written in this form, the state is considered separable; otherwise, it is entangled.

In quantum systems with three or more particles, there are primarily two interesting classes of entangled states: GHZ and W states. These two classes of entangled states can be partitioned into two disjoint categories through stochastic local operations and classical communication (SLOCC) \cite{li2007stochastic}. The GHZ state is a special type of multipartite entangled state in which the entanglement between all particles is global. In an $N$-particle system, the GHZ state can be expressed as:
\begin{equation}
    \left.|{\mathrm{\Psi}}_{GHZ}\right\rangle= \frac{1}{\sqrt2}\ (\left.|1\right\rangle^ {\otimes N}+\left.|0\right\rangle^ {\otimes N})
    \label{ghz}
\end{equation}

The W state is another type of multipart entangled state characterized by the fact that even if one particle is lost after the measurement, the remaining particles keep entangled, unlike the GHZ state, which becomes completely separable. In an $N$-particle system, the W state can be expressed as:

\begin{equation}
    \left.|{\mathrm{\Psi}}_W\right\rangle= \frac{1}{\sqrt N}\ \left.(|10\ldots0\right\rangle+\left.|010\ldots0\right\rangle+\ldots+\left.|0\ldots.01\right\rangle 
    \label{w}
\end{equation}
The W state is locally indistinguishable from other states in its equivalence class under local unitary transformations \cite{zhang2021strong}. SLOCC can distinguish between these two classes of entangled states because they exhibit different invariances under local operations and classical communication. This classification is significant for quantum information processing tasks, such as quantum computation and quantum communication, as it helps to understand the application and limitations of entanglement resources in these tasks.

\subsection{\label{sec:citeref}Entanglement Depth}
\begin{figure}[h]
    \centering
    \includegraphics[width=0.5\textwidth]{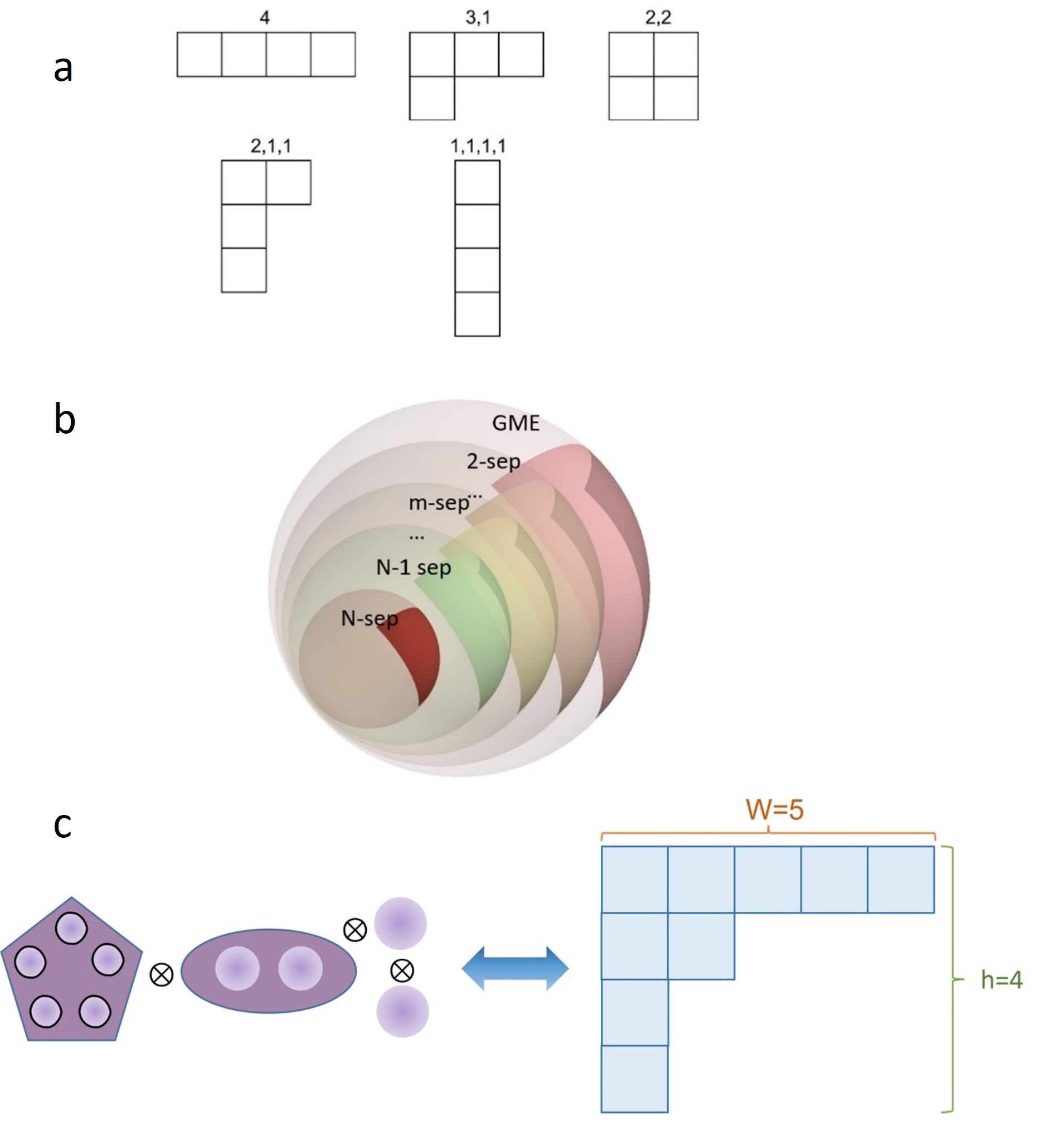}
    \caption{Yang diagram representation and entanglement Structure analysis of multi-body systems. (a) All possible Young diagrams for a 4-particle system;  (b) shows the separability hierarchy of N-qubit states. The m-separable(m-sep) state set $S_m$, where  $2 \leq m \leq N$, is the convex hull of separable states with different m-partitions. An N-qubit state is considered N-separable (N-sep) if it can be fully decomposed into N-independent quantum subsystems, implying that the entire quantum system exhibits no entanglement. Genuine multipartite entanglement (GME) \cite{toth2005detecting} occurs when a multi-qubit state cannot be represented as a convex combination of any separable states. In other words, GME represents an entanglement structure involving the entire multi-qubit system and cannot be described by decomposing it into smaller subsystems' entanglement structures. This indicates that the quantum state exhibits strong quantum correlations among all its subsystems. (c) characterizes the entanglement structure of multi-body systems using Young diagrams. In the depicted 9-body system, it can be divided into $h=4$ subsystems, with an entanglement depth of $w=5$, corresponding to the height $h$ and width $w$ of the Young diagram.}
    \label{fig1}
\end{figure}

An $N$-body system can be decomposed into multiple subsystems
$\Lambda=\{\Lambda_1, \Lambda_2, \ldots, \Lambda_k\},\ k\leq N$. This is a combinatorial problem. For an $N$-body system, there are $2^{N-1}$ types of partition, and $(N+1)$ kinds of split methods. Determining the exact analytic equation is very difficult, but recursive and dynamic programming methods can be used to calculate the integer division of $n$ particles.In general, we can use the Young diagram \cite{ren2021metrological} to directly show the partition. For example, a 4-body system has 8 types of partition and 5 types of split methods. This can be represented by the Young diagram shown in Fig.~\ref{fig1}(a).

The term $k$-producible represents the maximum degree of entanglement in the system. A $k$-producible pure state $\rho_{k-pro}$ can be written as 
\begin{equation}
  \rho_{k-pro}=\rho_{A1}\otimes\rho_{A2}\ldots\otimes\rho_{Am}
  \label{rou1}
\end{equation}
where $\rho_{Ai}$ denotes a state with a most maximum of $k$ particles. The maximum entanglement of a system corresponding to a $k$-producible pure state is $k$. A $k$-producible mixed state is defined as a convex combination of separable pure states with an entanglement depth of less than $k+1$. If a quantum state is not $k$-producible, this implies that its entanglement depth is $k+1$. We can also refer to the entanglement depth simply as $w$, which indicates that there are at least $w$ particles entangled together in the system.

The term $k$-separable represents the number of subsystems in a system. A $k$-separable pure state $\rho_{k-sep}$ can be expressed as $\rho_{k-sep}=\rho_{A1}\otimes\rho_{A2}\ldots\otimes\rho_{Ak}$, where the $N$-body system is divided into $k$ subsystems. A mixed state is called k-producible, if it is a mixture of pure states that are all k-producible at most. That is, the $k$-separable mixed state is a convex combination of separable pure states with more than $k-1$ subsystems. If a quantum state is not $k$-separable, it is considered $h$-inseparable.

The concept of $h$-inseparability implies that a system cannot be divided into $h$-separable subsystems. The larger the value of $w$ or the smaller the value of $h$, the more entanglement that exists in the multipartite system. This concept is intuitively illustrated in Fig.~\ref{fig1}(b)and(c).

\subsection{Dataset Preparing}
The training and testing datasets employed in this study are generated by following the above GHZ and W-state definition formulas, resulting in 200,000 density matrices, with 100,000 matrices corresponding to each of the GHZ and W states. To ensure a consistent distribution of data across entangled structures, density matrices are generated completely at random.

We consider a fixed number of particles and systematically generate states ranging from GME to fully separable, encompassing all possible cases. Taking the particle number as the known condition is by the real experimental situation, and thus would not damage the universality and effectiveness of our method. Since the situations arising are fixed and known, we randomized the probability coefficients of each subsystem to create a diverse dataset. To generate reliable data, we defined subsystems with several particles greater than or equal to 3 as the corresponding entangled states. For subsystems with two particles, we allowed the system to randomly select any Bell state. When the number of particles is equal to one, we let the system choose the zero state. Through this process, we generated the required for the experiments.

This approach to dataset generation ensures a comprehensive representation of entangled states while maintaining the rigor and objectivity necessary for academic research. By systematically exploring the full range of entangled structures and incorporating randomization into the probability coefficients, our dataset serves as a robust foundation for evaluating the performance of our machine learning and computer vision techniques for quantum entanglement detection.

\subsection{Hybrid CNN-Transformer model }
In our experiments, we pre-tested with a traditional fully connected neural network (FNN) \cite{sainath2015convolutional} and found that the results were excellent when the number of particles was small, but when the number of particles was greater than seven, the accuracy dropped significantly. This is because the FNN can not extract local features, it cannot capture the local spatial information of the input data. This problem can be solved by a CNN, which can effectively identify local features through a convolution operation, making it superior in image processing and other artificial intelligence fields.

 In essence, the FNN needs to flatten the input data into a one-dimensional vector, thus losing spatial information, whereas the CNN can accept the input data of the original shape, retain the spatial structure information, and help extract more efficient features. Through convolution operations, CNNs can efficiently extract local features while reducing the number of parameters. The addition of convolutional layers makes CNNs translational invariant and thus achieves remarkable success in image processing, computer vision, and other fields. However, CNNs still have limitations in terms of capturing long-distance dependencies.  With the development of computer vision technology, Vision Transformer(ViT) \cite{liu2021swin,dosovitskiy2020image}, which is the latest neural network development direction, aims to make full use of the advantages of Transformer structure. Transformers have made breakthroughs in the field of natural language processing, mainly because their self-attention mechanism can capture long-distance dependencies. ViT divides an image into small pieces and then processes these small pieces using a Transformer to capture global features. ViT has surpassed CNNs in many image recognition tasks and has become the current frontier technology in the field of computer vision. 

We adopted the perspective that quantum states can be regarded as analogous to images in the context of their mathematical representations. Drawing upon this conceptual similarity, we leverage advanced computer vision techniques to analyze these quantum states, effectively transforming the problem of entanglement detection into a task that parallels image analysis. This innovative approach facilitates the exploration of entanglement structure detection using well-established methods in the field of computer vision. 
\begin{figure}[h]
\centering
\includegraphics[width=0.35\textwidth]{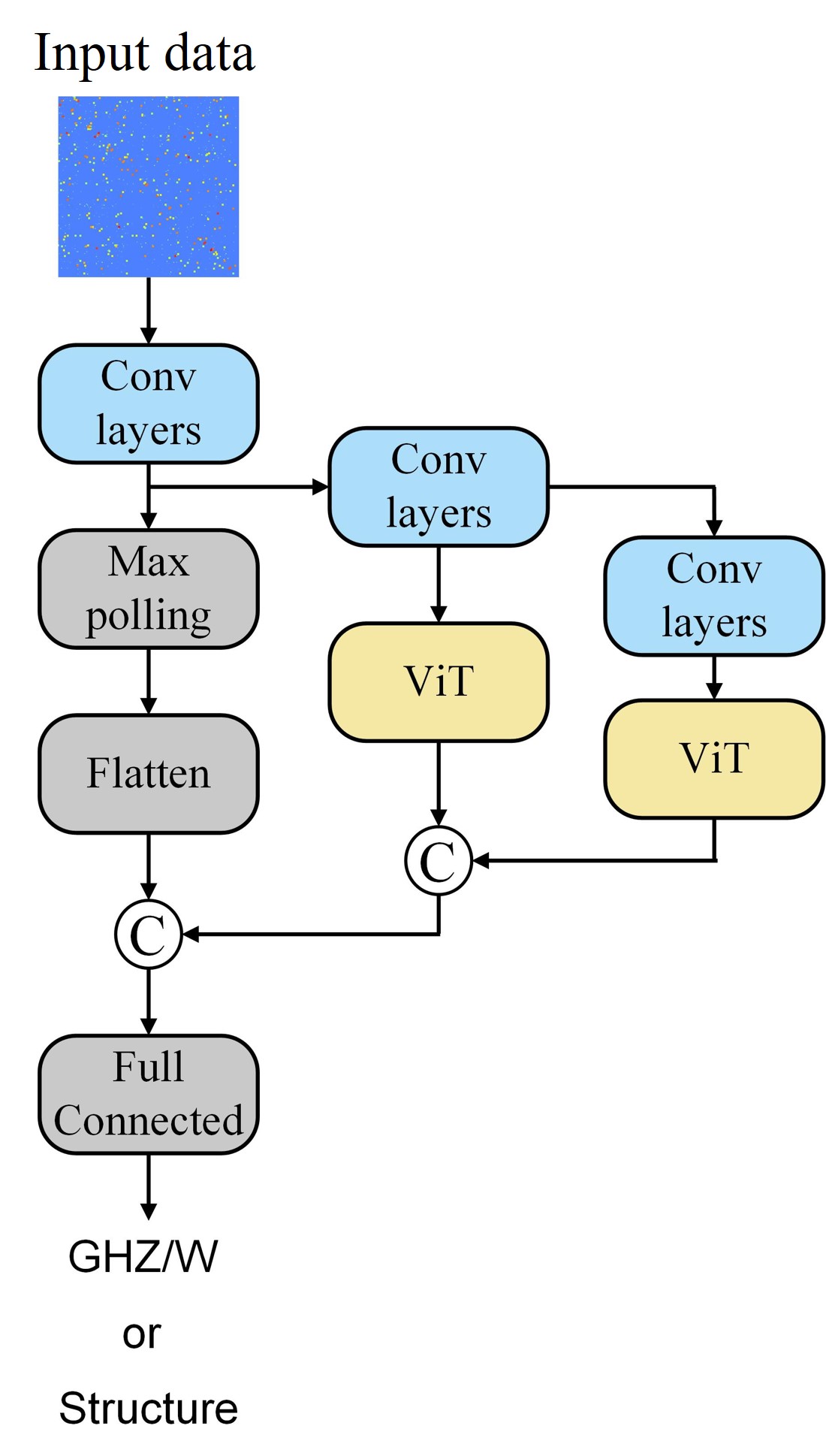}
\caption{Hybrid CNN-Transformer model }
\label{fig2}
\end{figure}
Inspired by the paper "An Image is Worth 16x16 Words: Transformers for Image Recognition at Scale" \cite{dosovitskiy2020image}, we employ CNNs and Transformers to extract local and global features, respectively. The dimensions of the density matrices vary with the number of particles. When the particle count is high, the density matrices become larger, necessitating more time and computational resources for the Transformer. To address this issue, we first applied a convolutional layer to extract features from the density matrices while altering their dimensions.

For particle numbers n=3, 4, 5, 6, and 7, the density matrix dimensions were relatively small, eliminating the need for a dimensionality reduction. However, for n = 8,9, and 10, the density matrix dimensions were larger, and we applied 2D convolutions to reduce their dimensions to 128x128. This process minimizes redundant features and decreases the computational load for the subsequent Transformer calculations. 

The network structures for different numbers of particles are shown in Fig.~\ref{fig2}. The convolutional layers (Conv layers) initially act as feature detectors, scanning the input data to highlight patterns that are akin to edges or textures in an image. These layers are critical for discerning local variations within the quantum data, which are essential clues to the nature of entanglement. The Max pooling layer serves to feature distill information by focusing on the most prominent features detected by the Conv layers, effectively reducing the dimensionality and computational complexity of the data. The Flatten operation then converts this condensed feature map into a one-dimensional array, setting the stage for a deeper analysis.

The ViT model includes a patch embedding process and position embedding. The patch embedding process divides these feature maps into smaller, manageable pieces or patches similar to breaking an image into segments. For different numbers of particles (n=3 to 10), we vary the patch sizes accordingly (from 2x2 to 32x32), optimizing the model's ability to process quantum states of varying complexity. Position embeddings are added to these patches to retain information about the relative or absolute position of the features within the original quantum state, which is crucial since the spatial relationship can hold significant quantum information.

The circle labeled 'C' in the figure represents the concatenation process. It combines the features extracted from separate pathways in a network. This step is essential for merging different types of information processed by the network—both local and global features into a comprehensive feature set.

Finally, the full-connected layers, a multilayer perceptron MLP \cite{gardner1998artificial}, take this richly processed information and determine the specific type of entanglement structure. This process mirrors the way we classify images based on a detailed understanding of their content, learned through both local and global observation.

\section{\label{sec:level3}Numerical Result}
In the following section, we present our numerical results including the evaluation metrics used to assess the performance of the model. By implementing the CNN and CNN-Transformer architectures, we showcase the successful classification of GHZ or W states and the detection of entanglement structures. Furthermore, we assessed the robustness of the model under various noisy settings, providing valuable insights for optimizing future quantum state classification tasks in real experiments.

\subsection{Classification of GHZ and W class state}
In our study, we initially focused on classifying GHZ and W states. We generated mixed states in the dataset, which comprised both GHZ and W states. Through the implementation of both CNN and CNN-Transformer models, we can accurately and effectively discern whether the quantum states of 3 to 11 particle systems contained GHZ or W states. By fine-tuning the model parameters, we attained a 100 \% accuracy rate in distinguishing between these quantum states.
\begin{figure*}
\centering
\includegraphics[width=0.9\textwidth]{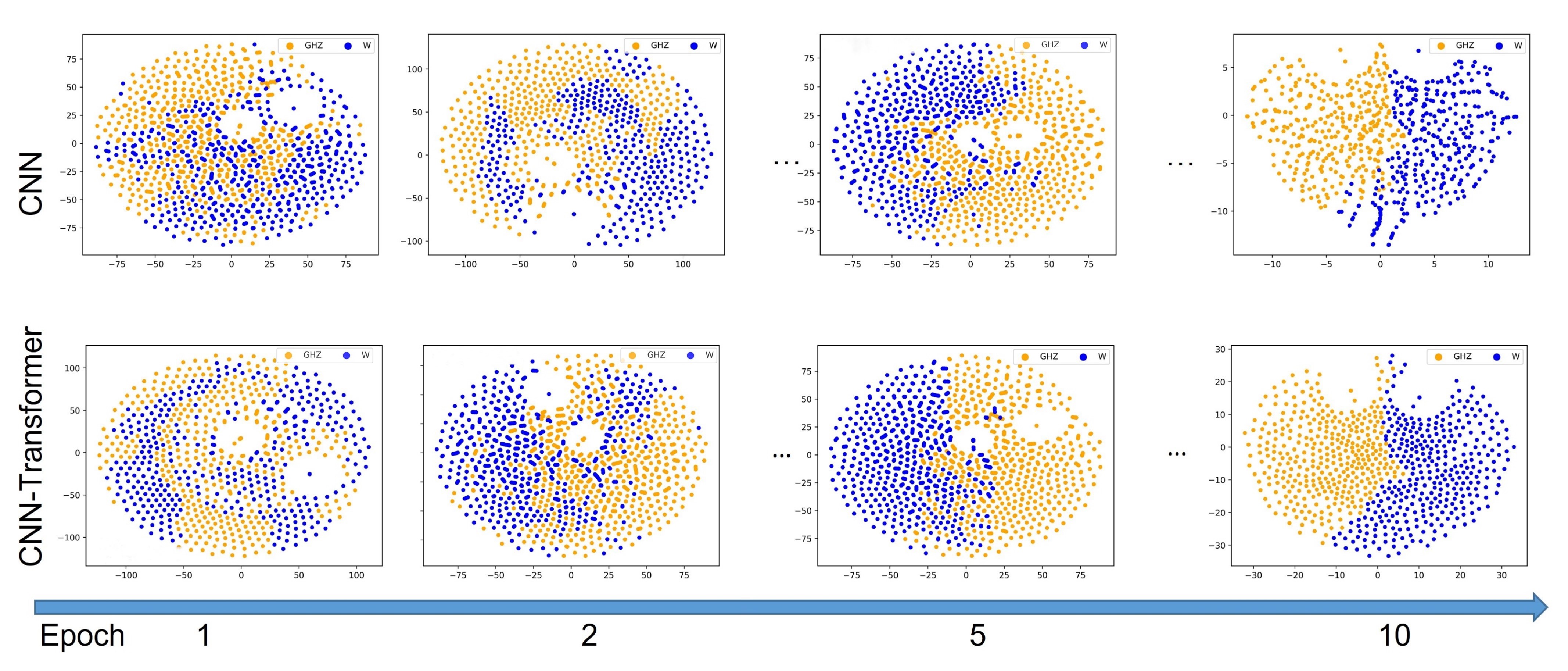}
\caption{During the training process, the evolution of feature vectors for 10,000 entangled samples of 5-qubit states.\label{fig3}}
\end{figure*}

We used t-distributed Stochastic Neighbor Embedding (t-SNE) \cite{belkina2019automated} plots to analyze our data. Fig.~\ref{fig3}. showed t-SNE plots for the case N=5. It can be seen that as training progressed the clustering features were more and more clearer. In early training, the GHZ and W state data points overlapped or were disorganized, indicating that the model had not fully learned their differences. In later epochs, the GHZ and W state data points separated and formed compact clusters, indicating that the model captured the high-dimensional data structure. This allows it to effectively differentiate between these two quantum states. As training continued, the clusters became more distinct, indicating improved classification performance.

Compared with the CNN model, the CNN-Transformer showed better classification on the t-SNE plot. It achieved clear separation and compactness of data point clusters in fewer epochs, learning distinguishing features more quickly. The CNN-Transformer model also had fewer outlier points, suggesting better handling of potential anomalies or noise. This difference in performance is due to the CNN-Transformer model combining CNN's local perception and Transformer's global perception, making it better at capturing complex data correlations and context. 

In summary, the CNN-Transformer model outperformed the CNN model when classifying the GHZ and W states. The t-SNE plots exhibited faster learning and better classification performance.

\subsection{Detecting entanglement structure}
Detecting entanglement structures, particularly in our massive dataset, is a challenging multi-classification problem. As the number of particles increases, the number of classifications exponentially increases, making the problem more complex. To address this issue effectively, we need to design a powerful model that can capture the entanglement features of different particle numbers.

As mentioned earlier, the CNN-Transformer model demonstrates superior performance in classifying the GHZ and W states. Its advantage lies in the combination of CNN's local perception capabilities and Transformer's global perception capabilities. This also allows the CNN-Transformer model to perform better in complex multi-classification problems. As the number of particles and classifications grows exponentially, we should further optimize the architecture of the CNN-Transformer model to maintain its high performance. Our approach involved increasing the number of layers of the model and adjusting its hyperparameters to enhance its expressive power.

\begin{figure*}
 \centering
\includegraphics[width=0.8\textwidth]{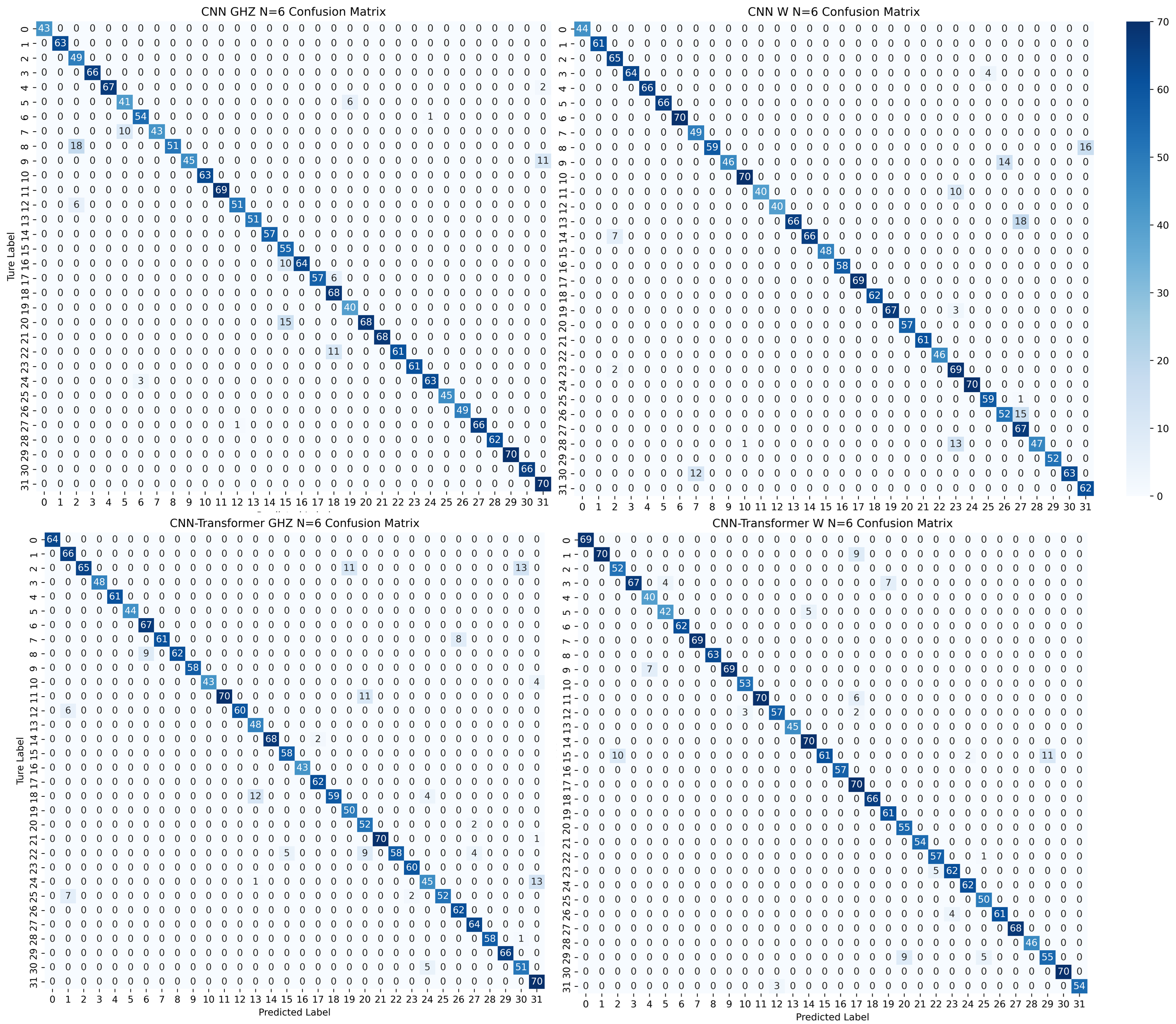}
\caption{The confusion matrices for six-particle entanglement structures. The confusion matrices for six-particle entanglement structures are displayed in the figure. In each confusion matrix, the horizontal axis (x-axis) represents the predicted class labels, while the vertical axis (y-axis) represents the true class labels. Each cell in the matrix contains a count of instances where the model predicted a particular class label (x-axis) for the instances of the true class label (y-axis). There are a total of 32 entanglement structures for six particles, resulting in label sequences from 0 to 31. The values in the cells are represented by varying shades of blue, with larger values corresponding to darker shades and smaller values appearing lighter, almost white. \label{fig4}}
\end{figure*}

To validate the performance of the CNN-Transformer model's performance in handling this multi-classification problem, we adopted confusion matrices \cite{visa2011confusion} as visualization tools. By examining the confusion matrices, we can observe the superior performance of the CNN-Transformer model in detecting entanglement structures of the GHZ and W states more intuitively. Fig.~\ref{fig4} shows the case N=6. The figure showcases four confusion matrices: the top-left corner represents the CNN model for the GHZ class with an accuracy of 92.25\%, the top-right corner represents the CNN model for the W class with an accuracy of 94.32\%, the bottom-left corner represents the CNN-Transformer model for the GHZ class with an accuracy of 94.25\%, and the bottom-right corner represents the CNN-Transformer model for the W class with an accuracy of 94.81\%. It can be noticed that the main diagonal elements (i.e. true positives and true negatives) of the CNN-Transformer model are larger than those of the CNN model, suggesting that the CNN-Transformer model has an advantage in predicting the correct number of GHZ and W state samples. This observation indicates that the CNN-Transformer model has higher accuracy in distinguishing these two quantum states. In addition, we find that the off-diagonal elements (i.e. false positives and false negatives) of the CNN-Transformer model are smaller compared to the CNN model.

This performance difference can be attributed to the CNN-Transformer model's combination of CNN's local perception capabilities and the Transformer's global perception capabilities, making it more advantageous in capturing correlations and contextual information in complex data. Therefore, when dealing with GHZ and W state classification tasks, the CNN-Transformer model achieves better performance than the pure CNN model.

As the particle number increases, the entanglement structures become more complex, and the density matrices exhibit higher dimensions, increasing the computational complexity. We anticipate that the accuracy and precision of our CNN and CNN-Transformer models may gradually decrease with an increase in the particle number. However, with complex optimization, the maximum particle number could potentially reach 15, while still achieving an accuracy of over 75\%. The scalability of our methods depends on the efficiency of feature extraction and the balance between local information preservation and computational complexity reduction. Further experiments are required to validate our results.

\subsection{\label{sec:citeref}Physical verification of Entanglement Detecting}

In practical applications, different types of noise make the precise determination of the entanglement structure more challenging. It is crucial to note that datasets formed by real experimental states may exhibit different distributions compared with random test datasets. This discrepancy is a primary challenge in real-world applications of machine learning, particularly when obtaining labels for test data is difficult or expensive, such as in bio-medicine \cite{mamoshina2016applications}, material science \cite{callister2007materials}, and physical science \cite{change2013physical} domains.

Fortunately, our hybrid CNN-Transformer model can address this issue. In the following, we test the performance of our model with real non-idealities in noisy intermediate-scale quantum (NISQ) devices \cite{preskill2018quantum}. OriginQ's quantum computing platform is favored for its accessibility and the maturity of its software development toolkit, Qpanda \cite{QpandaTutorial}. All subsequent physical experiments were conducted on a six-qubit superconducting quantum computer, OriginQ Wuyuan No.2 
 \cite{OriginQ}. Owing to the limitations of the current quantum computing capabilities, our data were restricted to three, four, and five particles.

\begin{figure*}
\centering
\includegraphics[width=0.85\textwidth]{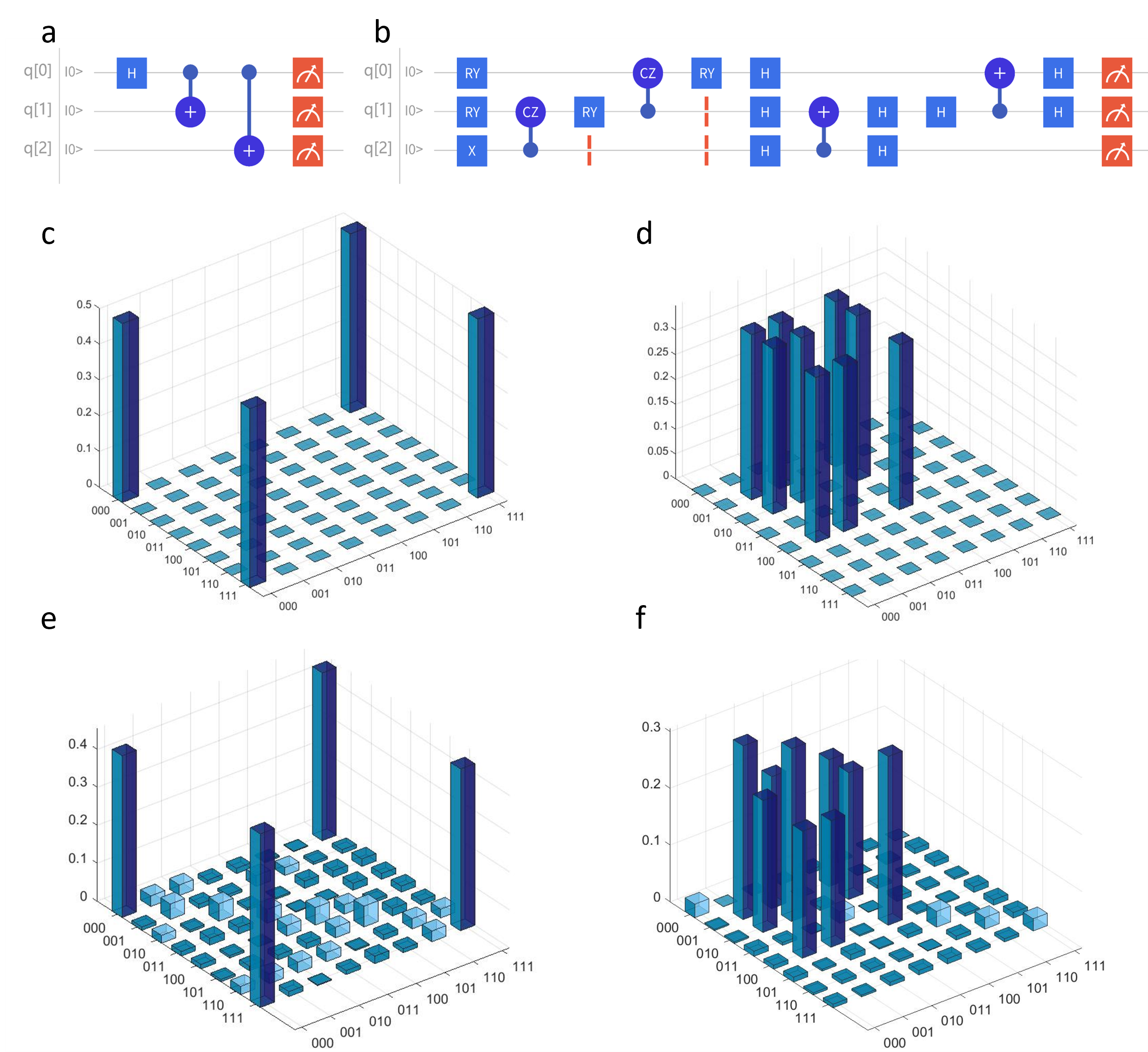}
\caption{(a) The circuit employed for generating the GHZ state on OriginQ hardware. The circuit is comprised of a sequence of operations executed on three qubits, namely q[0], q[1], and q[2]. The operations involve Hadamard gates (H) and controlled-NOT (CNOT) operations. (b) The circuit is utilized for generating the W state on OriginQ hardware. This circuit consists of a series of operations applied to three qubits, q[0], q[1], and q[2]. The operations include RY (Z-rotations), Hadamard gates (H), controlled-NOT (CNOT) operations, and controlled Z (CZ) gates. Moreover, barrier operations (BARRIER) are employed to ensure the proper execution order on the hardware. (c)(d) The ideal density matrices for the GHZ and W states, respectively. (e)(f) The experimentally obtained density matrices prepared using the OriginQ Wuyuan No.2 hardware, exhibited fidelities of 0.8187 and 0.7578 for the GHZ and W states, respectively. \label{fig5}}
\end{figure*}

A total of 4000 data were prepared as the test set in our pre-trained model. The dataset preparation is illustrated in Fig.~\ref{fig5}, considering the case of three particles as an example. Fig.~\ref{fig5}(a) and (b) show the quantum circuits for the generation of the 3-particle GHZ state and W state respectively. Fig.~\ref{fig5}(c) and (d) depict the density matrices numerically generated by the definitions of the GHZ and W state, whereas Fig.~\ref{fig5}(e) and (f) present the density matrices generated through the real quantum devices.

We first used the model to classify quantum states, achieving an experimental classification accuracy of 100\%. Subsequently, we classified the entanglement structures. The CNN-Transformer model achieved an accuracy of 99.27\% for three particles, 98.85\% for four particles, and 97.36\% for five particles. The experimental results demonstrated the reliability and robustness of our model.

However, it should be noted that in the NISQ device, as the number of particles increases, fidelity inevitably becomes lower and lower. In the quantum computer we used, fidelity drops to 0.6158 for the five particles. We noticed that the elements on the diagonal in the real quantum computer were significantly larger than those in other positions. Therefore, we added white noise to test the performance of our model more comprehensively. 

Quantum states with white noise were prepared for 6-10 particles.
\begin{equation}
   \rho_{ng} = p|\mathrm{\Psi}_{GHZ}\rangle\langle\mathrm{\Psi}_{GHZ}|+\frac{(1-p)I_n}{2^n},\quad p\in [0,1] 
   \label{noiseghz}
\end{equation}
\begin{equation}
   \rho_{nw} = p|\mathrm{\Psi}_W\rangle\langle\mathrm{\Psi}_W|+\frac{(1-p)I_n}{2^n},\quad p\in [0,1]
   \label{noiseww}
\end{equation}

{\noindent}where $|\mathrm{\Psi}_{GHZ}\rangle$ is as show in Eq.(\ref{ghz}),and$|\mathrm{\Psi}_W\rangle$ is as show in Eq.(\ref{w}).

\begin{figure}[h]
\centering
\includegraphics[width=0.49\textwidth]{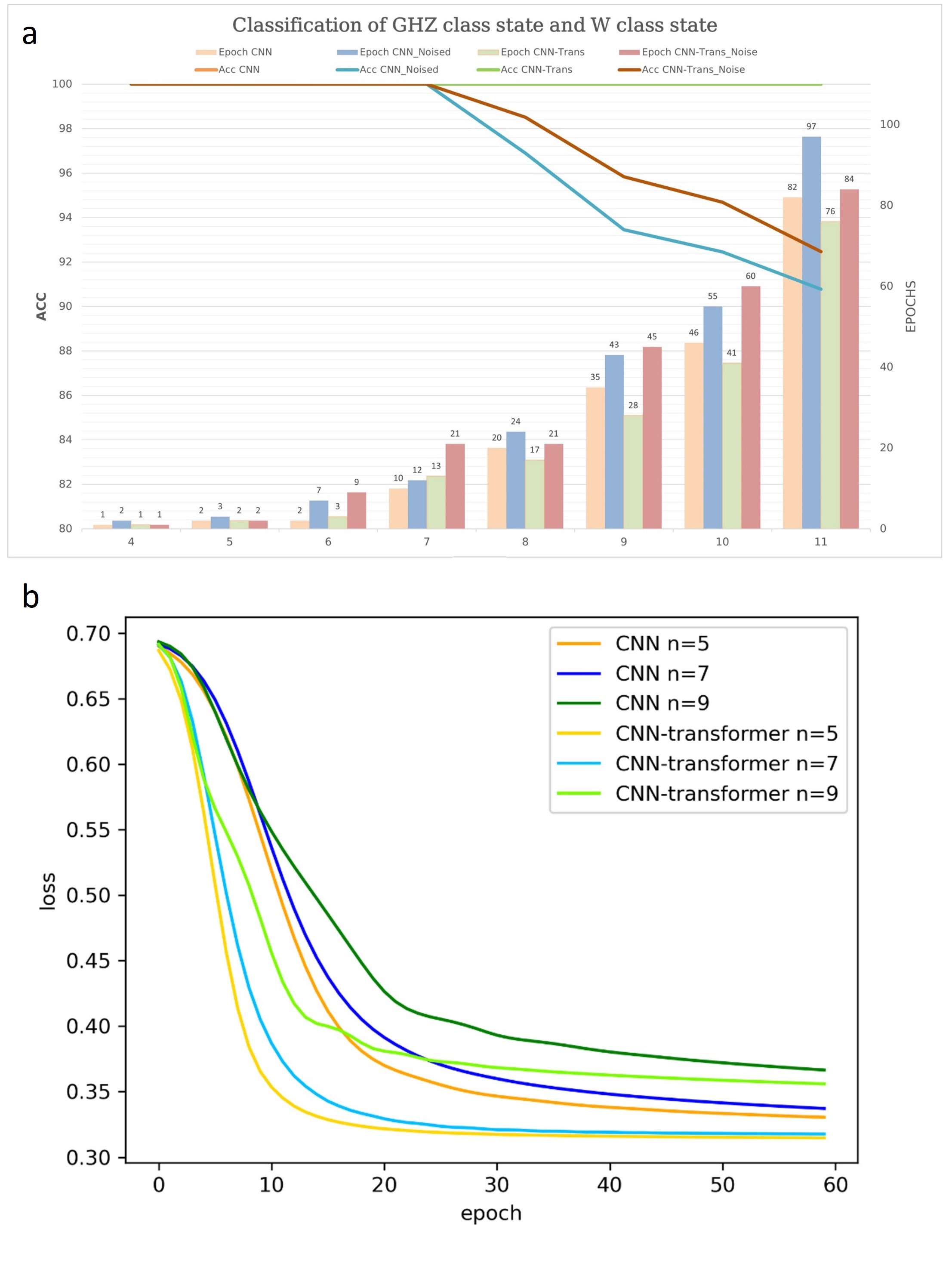}
\caption{(a) Classification of GHZ class state and W class state. The graph displays four distinct colors representing different models: orange for CNN, blue for CNN\_Noise, green for CNN-Transformer, and red for CNN-Transformer\_NOISE. The left y-axis indicates the accuracy of these four models at various particle numbers while the right y-axis denotes the number of iterations required to achieve high accuracy in machine learning training. (b)The plot illustrates the changes in the loss for both CNN and CNN-Transformer models during training with 5, 7, and 9 particles, as the number of epochs increases.}
\label{fig6}
\end{figure}

After generating the dataset, we first classified these states into GHZ and W classes. The classification performance is demonstrated from two perspectives, as shown in Fig.~\ref{fig6}(a). The first perspective is from accuracy, showing that accuracy remains at 100\% without noise even as the number of particles increases while in the presence of noise, relatively good results are maintained despite a slight decrease in detection accuracy. For 5 and 7 particles, both CNN and CNN-Transformer models achieve an accuracy of 100\%. However, as the number of particles increased to 9, the accuracy of the CNN model dropped to 93.45\%, while the CNN-Transformer model maintained a higher accuracy of 95.83\%. This observation demonstrates that despite the presence of noise, our models still perform well in handling complex systems with increasing particle numbers. Furthermore, the loss function values for training and validation converged to a low value, reinforcing the effectiveness and robustness of our models. The second perspective is from the required epoch, emphasizing that the exponential increase in data size with the increase in particle number leads to more training iterations. However, the growth of iterations remained linear, and the epoch termination condition employed early stopping to prevent model overfitting. Consequently, training cases automatically when there is no performance improvement. It is also noteworthy that, even as the particle number increases to 11, the number of training iterations remains below 100. This reflects a considerable level of efficiency and speed which is particularly important for large-scale systems.  We utilized ten-fold cross-validation to evaluate the reliability of our data.

Fig.~\ref{fig6}(b) displays the loss values of CNN and CNN-Transformer models during the training epochs for quantum systems with 
\( n=5, 7, \text{ and } 9 \). The curves demonstrate a steady decrease in loss, indicating learning and improvement in detecting entanglement structures as the training progresses. The CNN-Transformer models exhibit lower loss values compared to CNNs alone, especially in early epochs, suggesting faster convergence. Despite the increasing complexity with higher quantum bit numbers, CNN-Transformers maintain lower losses, highlighting their potential for handling larger quantum systems. The reduced number of epochs required for convergence also implies that our models can be trained more efficiently, achieving optimal performance in a shorter period.

\begin{figure}[h]
\centering
\includegraphics[width=0.45\textwidth]{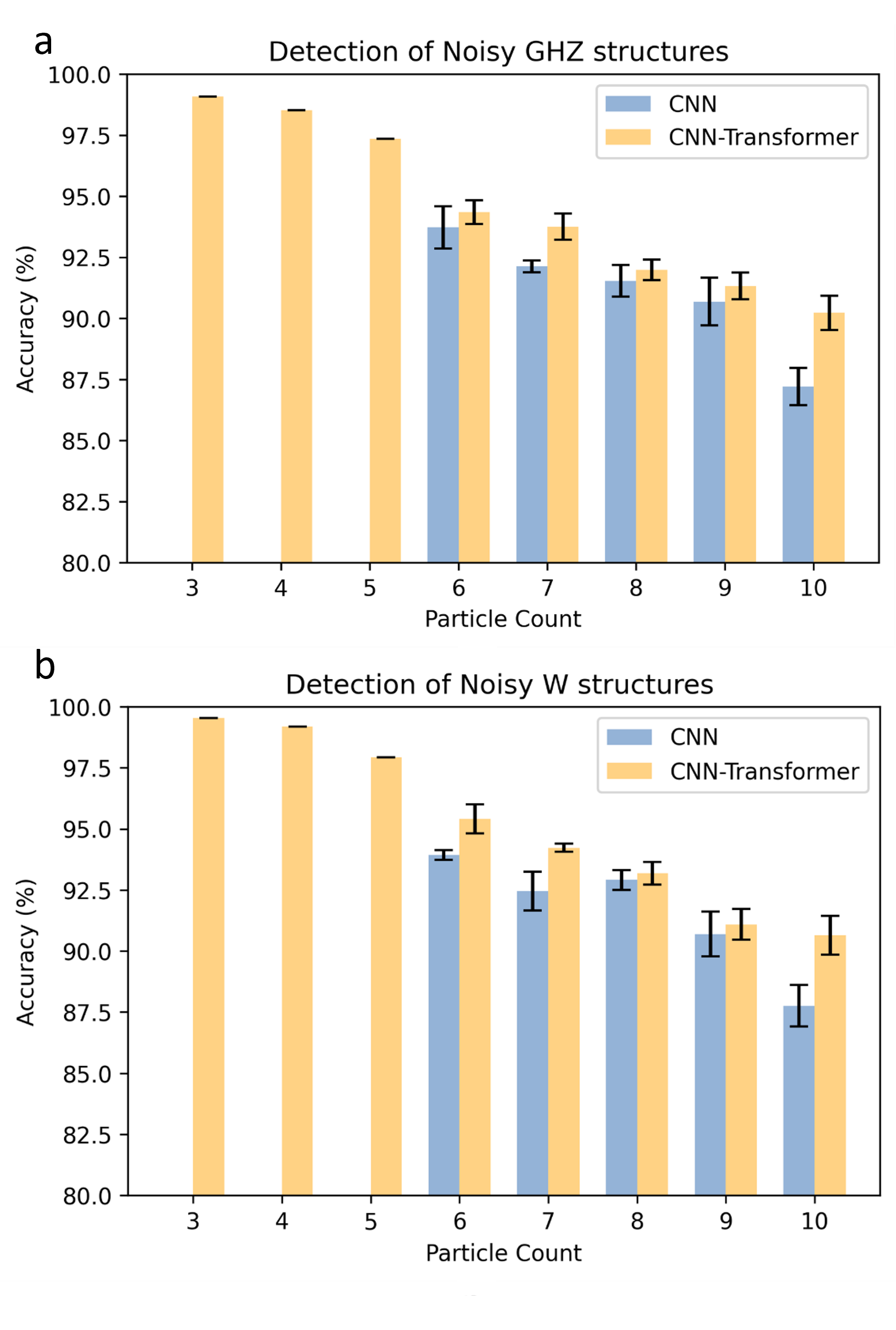}
\caption{(a) Shows the accuracy of detecting GHZ entanglement structures with noise.(b) Shows the accuracy of detecting W entanglement structures with noise.}
\label{fig7}
\end{figure}

Next, we use noisy data to detect entangled structures with particle numbers from 6 to 10. Fig.~\ref{fig7}(a) and (b) show the accuracies of the two models with noise. It is worth noting that our method achieves 95.42\% accuracy for 6 particles, and 90.23\% accuracy for 10 particles, which is slightly lower than the 93.52\% accuracy for the noise-free cases. Although the accuracy was somewhat affected, was still acceptable considering the small error bars. This indicates that our method exhibits good stability and robustness under different particle numbers and noise levels. To address the issue of decreased accuracy, we can construct more complex model structures, adjust the model hyperparameters, and adopt more advanced training strategies. In addition, we can enhance the ability of this model by increasing the diversity and size of the dataset to accomplish more tasks. This will help the model to perform better when dealing with various situations encountered in real-world applications.

\section{\label{sec:level4} Conclusion}
This article discusses the application of computer vision methods to identify the entanglement types and structures of quantum states simultaneously. Computer vision technologies, particularly the CNN and Transformer models, are inherently adept at handling and analyzing data in regular-sized matrices, quickly and accurately. Therefore, compared with other data-driven methods for entanglement detection, our hybrid CNN-Transformer model has the following advantages.

First, based on the same structure, our model can address a broader range of problems. In this work, we use the same trained neural network for both the classification of GHZ and W states and the detection of specific entanglement structures. If we expand the dataset, it also allows for the classification of Cat, Gaussian, and GKP states. Additionally, through detailed analysis of quantum states, the model can detect more entanglement structures, which is of significant importance in fields such as quantum communication and quantum computing. This demonstrates the great practicality of our model.

Second, our approach exhibits significant accuracy and efficacy. Numerical examples with 3-10 particles show that even with the presence of noise, our model achieves an average accuracy of 98.32\% for entanglement state classification and an average accuracy of 95\% for entanglement structure detection. Moreover, our model offers a remarkable balance between high performance and rapid data processing. The rapid decrease in the loss value demonstrates the effective convergence of our method.  Thus, our approach can effectively reduce time and computational costs and has the potential for applications to large-scale systems.

In conclusion, benefiting from the powerful data processing capabilities of computer vision, we obtain high accuracy for both classification of entangled states and detection of various entanglement structures with low time and computational cost. Here, we just consider the cases of 3-10 particles because of the limitation of the current Quantum State Tomography (QST) \cite{lvovsky2009continuous,palmieri2020experimental} rather than the capability of our model. In the next work \cite{li2024}, we adopt an innovative approach to overcome this limitation and apply our model to systems with a large number of qubits.

%As is known to all, experimentally certifying and quantifying states of quantum systems are too complex for QST to be a feasible option, which is typically confined to handling up to 10 particles. Even with full information about the quantum state, determining its entanglement structure remains an NP-hard problem \cite{ma2018transforming}. This is mainly because the density matrix is not an ideal representation of quantum states for entanglement detection. To overcome this limitation, we adopt an innovative approach , utilizing computer vision technology, particularly the hybrid CNN-Transformer model, to extend the capability of quantum state detection. The essence of this method is to transform complex quantum measurement data into a format that computer vision models can comprehend and process, thereby facilitating effective detection of higher particle number entanglement structures. Our preliminary experimental results demonstrated the feasibility of this method for enhancing the detection of entanglement structures across a higher number of particles. This not only showcases the potential application of computer vision technology in the field of quantum information science but also opens new avenues for future quantum experiments and state analysis.

\begin{acknowledgments}
This work is supported by the National Natural Science Foundation of China under Grants No.61975005, Beijing Academy of Quantum Information Science under Grants No.Y18G28 and the Fundamental Research Funds for the Central Universities under Grants No.YWF-22-L-938.
\end{acknowledgments}

\nocite{*}

\bibliography{apssamp}% Produces the bibliography via BibTeX.

\end{document}